\pdfoutput=1

\documentclass[11pt]{article}

\usepackage[final]{acl}

\usepackage{times}
\usepackage{latexsym}

\usepackage[T1]{fontenc}

\usepackage[utf8]{inputenc}

\usepackage{microtype}

\usepackage{inconsolata}

\usepackage{graphicx}

\title{LLM-Driven Learning Analytics Dashboard for Teachers\\ in EFL Writing Education}

\author{
    Minsun Kim, SeonGyeom Kim, Suyoun Lee, Yoosang Yoon,\\
    \textbf{Junho Myung, Haneul Yoo, Hyunseung Lim, Jieun Han, Yoonsu Kim, } \\
    \textbf{So-Yeon Ahn, Juho Kim, Alice Oh, Hwajung Hong, }\textbf{Tak Yeon Lee\thanks{Corresponding author.}} \\
    KAIST, South Korea \\
    \texttt{\{\href{mailto:9909cindy@kaist.ac.kr}{\color{black}{9909cindy}}, 
    \href{mailto:sgk_0320@kaist.ac.kr}{\color{black}{sgk\_0320}}, 
    \href{mailto:jenslee705@kaist.ac.kr}{\color{black}{jenslee705}}, 
    \href{mailto:soulmilk98@kaist.ac.kr}{\color{black}{soulmilk98}}, 
    \href{mailto:junho00211@kaist.ac.kr}{\color{black}{junho00211}}, 
    \href{mailto:haneul.yoo@kaist.ac.kr}{\color{black}{haneul.yoo}}, }\\
    \texttt{
    \href{mailto:charlie9807@kaist.ac.kr}{\color{black}{charlie9807}},
    \href{mailto:jieun_han@kaist.ac.kr}{\color{black}{jieun\_han}},
    \href{mailto:yoonsu16@kaist.ac.kr}{\color{black}{yoonsu16}},
    \href{mailto:ahnsoyeon@kaist.ac.kr}{\color{black}{ahnsoyeon}},}
    \texttt{
    \href{mailto:juhokim@kaist.ac.kr}{\color{black}{juhokim}}\}@kaist.ac.kr},\\
    \texttt{\href{mailto:alice.oh@kaist.edu}{\color{black}{alice.oh@kaist.edu}}},
    \texttt{\{\href{mailto:hwajung@kaist.ac.kr}{\color{black}{hwajung}},
    \href{mailto:takyeonlee@kaist.ac.kr}{\color{black}{takyeonlee}}\}@kaist.ac.kr}\\ 
}

\begin{document}
\maketitle
\begin{abstract}
This paper presents the development of a dashboard designed specifically for teachers in English as a Foreign Language (EFL) writing education. Leveraging LLMs, the dashboard facilitates the analysis of student interactions with an essay writing system, which integrates ChatGPT for real-time feedback. The dashboard aids teachers in monitoring student behavior, identifying noneducational interaction with ChatGPT, and aligning instructional strategies with learning objectives. By combining insights from NLP and Human-Computer Interaction (HCI), this study demonstrates how a human-centered approach can enhance the effectiveness of teacher dashboards, particularly in ChatGPT-integrated learning.
\end{abstract}

\section{Introduction}
The creation of GPT\citep{brown2020language}, an LLM developed by OpenAI, has catalyzed groundbreaking changes across various sectors of society. It has emerged as a transformative tool in education \citep{kasneci2023chatgpt}, particularly in the domain of language learning \citep{hong2023impact}. Specifically, many applications utilizing GPT models have been adopted in English as a Foreign Language (EFL) contexts \citep{nasrullah2024application}, such as generating personalized feedback for students and creating learning resources \citep{han2023fabric, amin2023ai}. However, while many educational tools have been developed for students that possess the potential to replace teachers \citep{koraishi2023teaching}, there has been a significant lack of tools developed specifically for teachers \citep{ausat2023can}. Additionally, there is a noticeable scarcity of education applications utilizing fine-tuned models or datasets made in real-world settings \citep{grassini2023shaping}.

This study presents how an NLP-integrated dashboard can be applied to enhance the teaching process in real-world educational contexts. This dashboard analyzes data from RECIPE, a student-side system used in EFL writing education \citep{han2023recipe}. The RECIPE system integrates ChatGPT to help students improve their writing by providing real-time feedback and scoring. The dashboard helps teachers monitor and evaluate student interactions with RECIPE, offering insights into student behavior, identifying noneducational interactions, and supporting the alignment of learning objectives with student activities. 

The main contribution is as follows:
\begin{enumerate}
    \item We introduce a novel dashboard application that leverages NLP models specifically designed to support teachers in ChatGPT-integrated EFL writing education contexts.
    \item By combining insights from HCI with NLP, this study demonstrates a specific methodology for how a human-centered learning analytics approach can enhance the applicability of LLM in educational settings.
\end{enumerate}

\section{Related Work}
\subsection{LLM applications for teachers}
Discussions among teachers often focus on the positive and negative impacts of LLMs on student learning \cite{steele2023gpt}, yet there is a noticeable lack of tools directly aimed at assisting teachers. While some tools are being developed to support teachers' learning \citep{markel2023gpteach}, others are typically limited to tasks such as automatic scoring or generating feedback for students \citep{han2023fabric, latif2024fine}. However, in traditional education, substantial research has been conducted in the field of learning analytics \citep{jivet2018license, matcha2019systematic}, particularly for process evaluation and overall student management \citep{wang2021applying, park2015development}. This paper aims to fill the gap by providing a tool specifically designed to support teachers in their evaluation and student management tasks.

\subsection{Human-centered NLP (HCNLP)}
The discourse around HCNLP has traditionally focused on incorporating human evaluations into NLP model development or creating models that consider human factors in their reasoning processes \citep{lynn2017human, das2023state}. However, in fields like HCI, Human-centered AI (HCAI) is increasingly being used as a term that encompasses the entire process—from collecting human needs to designing models and developing applications based on those needs \citep{hollund2022human, skeen2022integrating, auernhammer2020human}. Despite this broader approach, collaborations between NLP experts and HCI specialists to develop real-world applications have been, to our best knowledge, extremely rare \citep{holzinger2013combining}. This paper aims to demonstrate the impact of collaboration and user input during the NLP model development process.

\subsection{Human-Centered Learning Analytics (HCLA)}
The concept of HCLA was introduced, focusing on incorporating human elements into learning analytics dashboards (LADs) design \citep{ahn2019designing}. Previous LAD design methods primarily aimed to enhance usability, reduce time spent skimming data, and support accurate interpretation of information \citep{verbert2013learning, kaliisa2023cada}. However, rooted in human-computer interaction, HCLA emphasizes designing technology that centers on human values and integrates user participation through co-design methods \citep{buckingham2019human}. Also, it is essential to account for the involvement of educational stakeholders who may lack expertise in data and analytics \citep{martinez2023human}. In this paper, we adopt HCLA principles to design a human-centered LAD through an iterative process, focusing on qualitative insights from a dashboard prototype using real-world classroom datasets.

\section{Methodology for Human-Centered NLP}
\subsection{Participants}
A team comprising NLP and HCI researchers was formed to develop this dashboard. Additionally, a TESOL\footnote{Teaching English to Speakers of Other Languages.} expert was involved to ensure the tool's effectiveness in analyzing English language learners. Before designing the dashboard, student data collection was conducted \citep{han2023recipe}. Our experiment is an empirical study within a real-world setting, we collected student data by recruiting students who attend an EFL writing course in the university. The student data involves timestamps, student essay editing logs, ChatGPT interaction logs, and satisfaction of ChatGPT responses. Six university EFL faculty members were recruited and consulted to understand the specific needs for the dashboard design. This collaborative approach was essential in ensuring that the dashboard meets the practical needs of teachers while being grounded in both NLP and HCI expertise.

\subsection{Progress}
\begin{figure}[t]
\centering
  \includegraphics[width=\columnwidth]{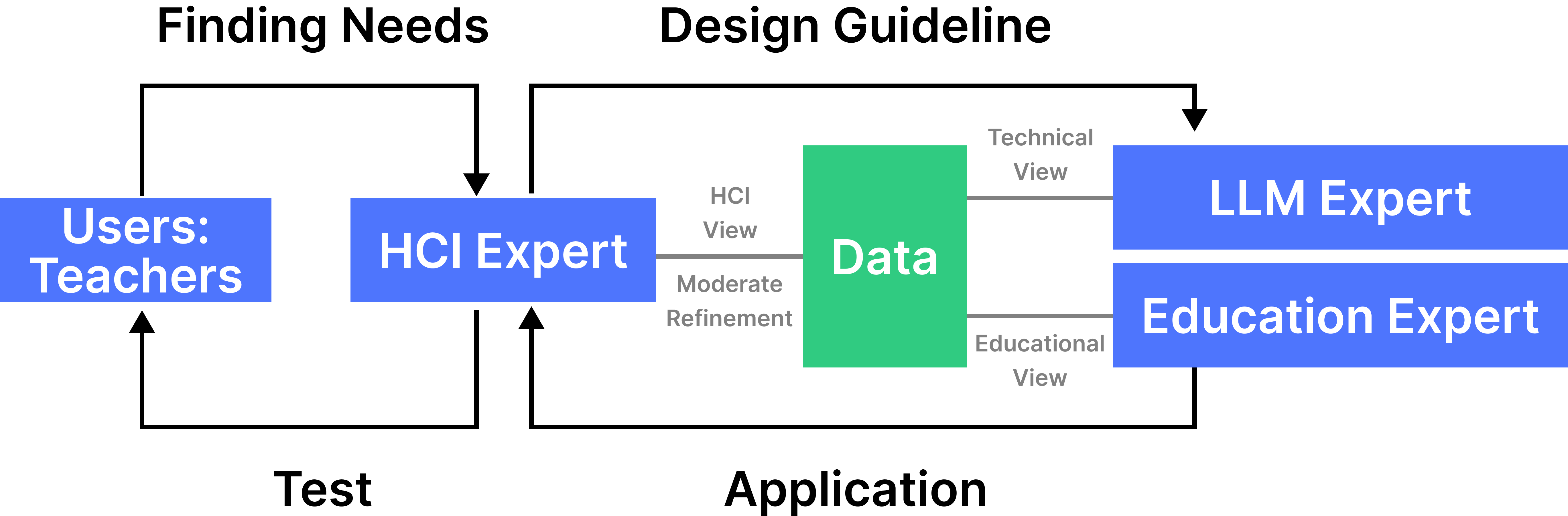} \hfill
  \caption{Relationships and progress between experts and potential users in education.}
  \label{fig:roles}
\end{figure}

The overall progress is shown in Figure \ref{fig:roles}. The dashboard design process consisted of two rounds of interviews. To ensure the design was grounded in real teaching experiences, we first conducted interviews with six teachers. These interviews were led by HCI researchers, who organized the specific design needs and goals into a comprehensive design guideline. The first phase of the study aimed to gather two types of findings: (1) teachers' perceptions of using ChatGPT in English education, and (2) initial design inspirations for the dashboard by observing how teachers track, evaluate, and guide student-ChatGPT interactions. Based on the insights gathered, we developed a prototype of the dashboard for teachers before the second phase of the study. The prototype features a wide range of charts and other components, displaying actual essays and log data collected throughout the semester. 

Following this, all researchers collaborated to identify how to elaborate data for the dashboard. To analyze the student chat log data, we applied NLP models. Since this process was tailored to the context of using ChatGPT in a university EFL writing curriculum, new data collection and refinement were necessary, including aligning the analysis with the learning objectives of university textbooks. After developing the initial dashboard prototype, we conducted interviews again with previous participants. To collect further feedback on the prototype and explore its potential impact in future EFL classes, we introduced the prototype, allowing teachers to try out its various components.

Throughout the study, we collected voice recordings of the interviews. While transcribing these interviews, we also made notes based on our observations. After conducting a qualitative analysis of this data, we derived design implications by synthesizing insights from both rounds of interviews.

\section{Dashboard Prototype}

\subsection{Dataset and models}
\begin{figure*}[t]
    \centering
  \includegraphics[width=0.9\linewidth]{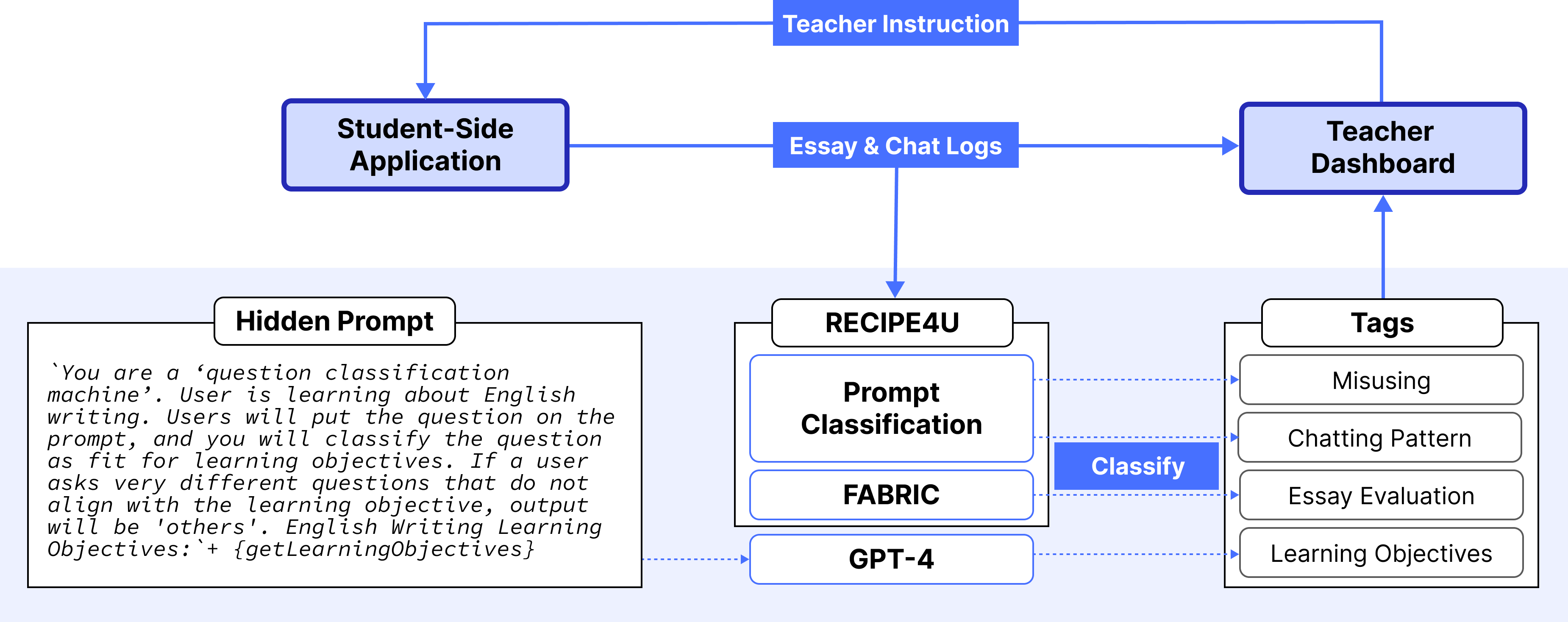} \hfill
  \caption {A dashboard system architecture.}
  \label{fig:architecture}
\end{figure*}

The dashboard is part of a larger system, as outlined in Figure \ref{fig:architecture}, where students use student-side application \citep{han2023recipe} to write essays every week and interact with ChatGPT. The essays and chat logs are automatically analyzed using three LLM models.
To detect and highlight AI misuse and chatting patterns, we employed fine-tuned GPT-4 and identified misuse by analyzing patterns of interaction between students and ChatGPT. Specifically, we defined misuse based on the most frequently occurring patterns among the twelve identified by the model. These patterns include requests for entire essay generation, excessive use of paraphrasing tools, and non-academic inquiries.
FABRIC contains an automated essay-scoring (AES) model based on teachers' essay grading data \citep{han2023fabric}. This is rubric-based AES using BERT \citep{devlin2018bert}. The pattern identification results and grading data are combined into the RECIPE4U dataset, and utilized in the overall teacher dashboard\citep{han2024recipe4u}. Also, \ref{fig:architecture} illustrates GPT-4 categorizes student prompts into learning objectives. This process is simply conducted with prompt engineering without evaluation. Learning objectives are from the university EFL lecture book.

\subsection{Feature description} 
\begin{figure*}[t]
  \includegraphics[width=\linewidth]{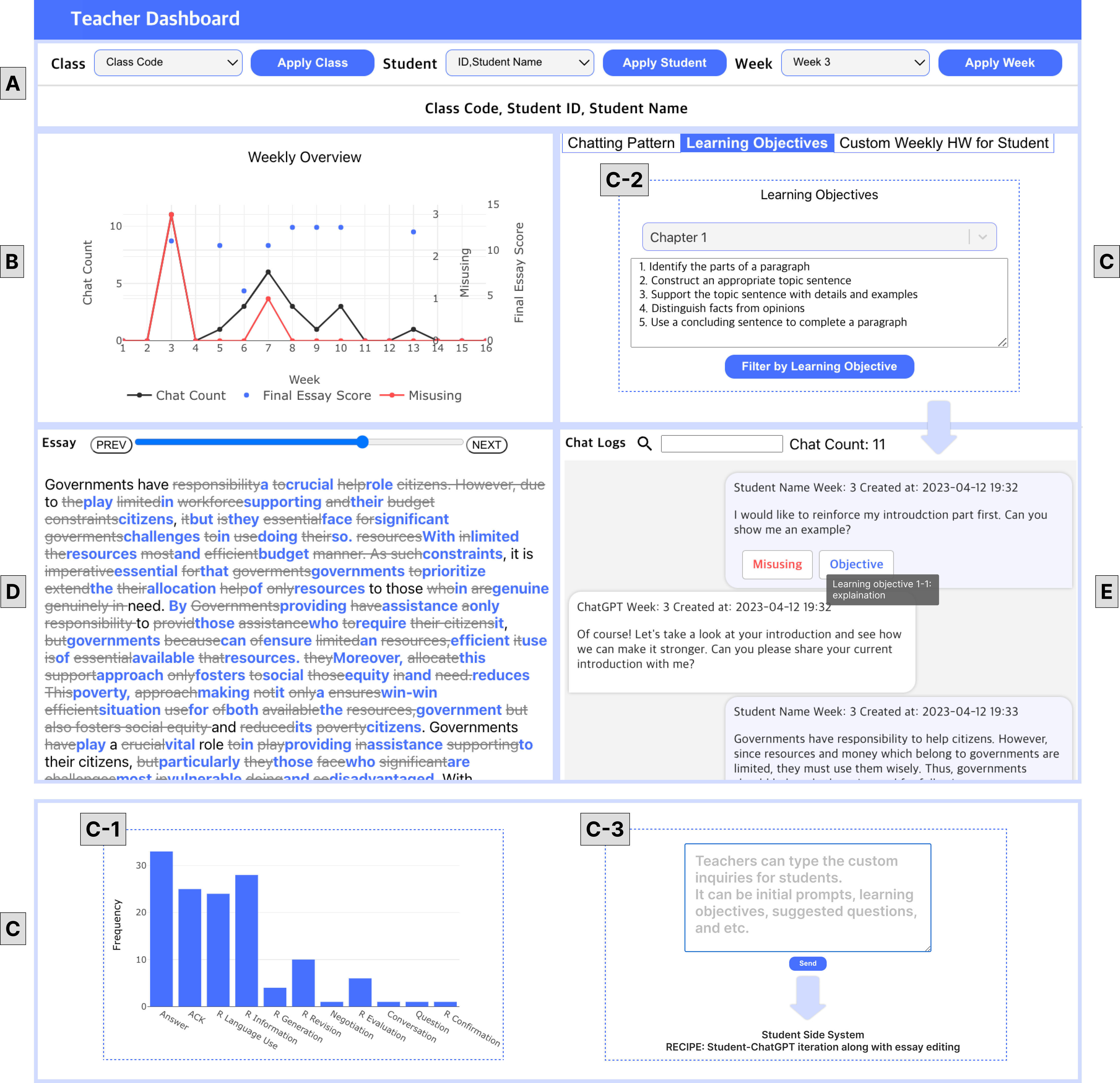} \hfill
  \caption {A dashboard prototype overview.}
  \label{fig:prototype}
\end{figure*}

We aimed to address specific needs and challenges from teacher interviews associated with integrating ChatGPT into educational settings. Each design goal(DG) is outlined with detailed descriptions, associated challenges, derived needs, and implemented features (Figure \ref{fig:prototype}).

\begin{itemize}
\setlength\itemsep{1em}
\item \textbf{DG1. Providing a quick overview of student-ChatGPT interaction. }
    \begin{itemize}
        \item \textbf{Description:} The dashboard offers various charts that give teachers a quick summary of student interactions with ChatGPT.
        \item \textbf{Challenges:} Teachers need an efficient way to monitor multiple students' interactions without being overwhelmed by data.
        \item \textbf{Needs:} Tools to visualize chat frequency, essay evaluations, and misuse instances clearly and concisely.
        \item \textbf{Features:} Weekly chat frequency charts, essay evaluation summaries, and misuse frequency charts (A). It also visualizes 12 common patterns of Student-ChatGPT dialogue \citep{han2023exploring} (C-1).
    \end{itemize}

\item \textbf{DG2. Identifying common patterns in undesirable usage of ChatGPT. }
    \begin{itemize}
        \item \textbf{Description:} The dashboard highlights AI-detected misuse or non-educational prompts.
        \item \textbf{Challenges:} Preventing students from misusing ChatGPT for non-educational purposes.
        \item \textbf{Needs:} Effective identification and reporting of undesirable usage patterns.
        \item \textbf{Features:} Highlighting and tagging of non-educational prompts, automated detection of misuse patterns (E).
    \end{itemize}

\item \textbf{DG3. Facilitating in-depth analysis of essay editing and chat log. }
    \begin{itemize}
        \item \textbf{Description:} The dashboard helps teachers understand students' behavior by displaying essay editing history alongside the chat log.
        \item \textbf{Challenges:} Teachers need comprehensive insights into student behavior to tailor their instructional strategies.
        \item \textbf{Needs:} Integrated views of editing history and chat logs, ability to filter by learning objectives.
        \item \textbf{Features:} Displaying essay editing history and corresponding chat logs (D), (E), listing learning objectives for each class unit, and filtering prompts by specific objectives (C-2), (E).
    \end{itemize}

\item \textbf{DG4. Customizing Prompt Instructions.}
    \begin{itemize}
        \item \textbf{Description:} The dashboard allows teachers to deliver personalized instructions to students.
        \item \textbf{Challenges:} Ensuring that ChatGPT provides responses aligned with the educational goals and learning objectives.
        \item \textbf{Needs:} Mechanisms to customize ChatGPT instructions based on teacher input.
        \item \textbf{Features:} Feature to add custom instructions that are transmitted to the student-side system, guiding students according to specific educational goals (C-3).
    \end{itemize}
\end{itemize}

\section{Discussion}
Through interviews and thematic analysis, we explored the role of dashboards in English education and the integration of ChatGPT into classrooms.

First, dashboards should support contextual analysis of student behavior. While current dashboards rely on visual and statistical analytics, our findings show that teachers prefer tools that facilitate semantic analysis, helping them understand the rationale behind students' prompts and responses. Unfortunately, reviewing each chat log line by line posed a significant challenge for teachers. They proposed several ideas to facilitate contextual analysis, such as employing LLM or other advanced AI technology to generate summaries of student behaviors and decipher the meanings behind them.

Additionally, dashboards aid teachers in designing adaptive learning loops. Beyond knowledge transfer, teachers craft curricula to optimize learning. Teachers expressed interest in integrating ChatGPT but raised concerns about potential misuse, such as ethical violations or inaccuracies. However, teachers may have different opinions of whether a certain use case is either educationally beneficial or detrimental, depending on educational contexts and teaching philosophy. Therefore, the dashboard must allow teachers to tailor their learning environments, including strategies for detecting and mitigating misuse. 

Lastly, dashboards should foster self-reflection on teaching practices. Educational dashboards also serve as tools for teachers’ professional development, enabling reflection on their methods. In the context of ChatGPT-integrated learning, most teachers would continuously monitor common patterns of misusing prompts, and develop their strategies for mitigating them. They also would like to evolve personalized guidance, such as recommending sample prompts to struggling students. These ideas illustrate the teachers' strong motivation to enhance their professional practices; nevertheless, the current version of both ChatGPT and dashboards has ample room for improvement. 

To successfully implement LLM in educational environments, collaboration between LLM experts, education experts, and HCI researchers is essential. LLM experts focus on the technical aspects, such as developing data classification models and applications, ensuring the models are robust, efficient, and capable of handling complex educational content. Education experts play a crucial role in evaluating and adjusting these applications to meet pedagogical standards and enhance the learning experience. HCI principles guide the refinement of these applications to improve user experiences. By considering these aspects, we can create innovative and effective educational tools that enhance student learning outcomes through the use of LLM.

\section{Conclusion}
The study introduces a novel teacher-focused dashboard that integrates advanced NLP techniques with human-centered design principles to improve the management of student learning processes in EFL contexts. By providing real-time analytics on student-ChatGPT interactions and essay revisions, the dashboard supports teachers in evaluating and guiding student performance more effectively. Initial feedback from educators underscores the potential of this tool to bridge the gap between AI advancements and practical teaching needs. Future iterations will aim to refine the dashboard through ongoing user feedback, ensuring its continued relevance and effectiveness in educational environments.

\section*{Acknowledgments}
This work was mainly supported by Elice\footnote{https://elice.io/en}, a leading company in the domain of digital education. The Azure credits for hosting the ChatGPT service were supported by the Microsoft Accelerate Foundation Models Research (AFMR) grant program\footnote{https://www.microsoft.com/en-us/research/collaboration/accelerating-foundation-models-research/}.

\section*{Ethical Statement}
All studies in this research project were performed under our institutional review board (IRB) approval. There was no discrimination when recruiting and selecting EFL teachers regarding any demographics, including gender and age. We set the wage per session to be above the minimum wage in the Republic of Korea in 2023 (KRW 9,260; USD 7.25). They were free to participate in or drop out of the experiment, and their decision did not affect their professional assessments. We deeply considered the potential risk associated with showing essays written by students in terms of privacy and personal information, and thus filtered out all sensitive information related to their privacy and personal information. 

\bibliography{acl_latex}




\end{document}